\documentclass{aa}

\usepackage{graphicx}
\usepackage{rotating}
\usepackage{txfonts}

\begin{document}

\title{Large Amplitude Photometric Variability of the Candidate Protoplanet TMR-1C}

\author{B. Riaz
              \inst{1}
               \& E. L. Mart\'{i}n
              \inst{2,}\inst{1}}

\institute{Instituto de Astrofis\'{i}ca de Canarias, E-38200 La Laguna, Tenerife, Spain\\ 
                \email{basmah@iac.es; ege@iac.es} \and Centro de Astrobiolog'a (CSIC/INTA), 28850 Torrej\'{o}n de Ardoz, Madrid, Spain}

\date{Recieved --- ; Accepted ---}

\abstract{}{In their {\it HST}/NICMOS observations, Terebey et al. 1998 (T98) detected a candidate protoplanet, TMR-1C, that lies at a separation of about 10$\arcsec$ ($\sim$1000 AU) from the Class I protobinary TMR-1 (IRAS 04361+2547) located in the Taurus molecular cloud. A narrow filament-like structure was observed extending south-east from the central proto-binary system towards TMR-1C, suggesting a morphology in which the candidate protoplanet may have been ejected from the TMR-1 system. Follow-up low-resolution spectroscopy by Terebey et al. 2000, however, could not confirm if this object is a protoplanet or a low-luminosity background star.}
{We present two epochs of near-infrared photometric observations obtained at the CFHT of TMR-1C. The time span of $\sim$7 years between the two sets of observations provides with an opportunity to, (a) check for any photometric variability similar to that observed among young stellar objects, which would indicate the youth of this source, and, (b) determine the proper motion.} 
{TMR-1C displays large photometric variability between 1 and 2 mag in both the {\it H}- and $K_{s}$-bands. From our 2002 observations, we find a ($H-K_{s}$) color of 0.3 mag, which is much bluer than the value of 1.3 mag reported by T98 from HST observations. Also, we observe brightening in both the {\it H}- and $K_{s}$-bands when the colors are bluer, i.e. the object gets redder as it becomes fainter. We have explored the possible origins for the observed variability, and find extinction due to the presence of circumstellar material to be the most likely scenario. The observed large-amplitude photometric variations, and the possible presence of a circumstellar disk, are strong arguments against this object being an old background star. }{} 


\maketitle

\section{Introduction}

TMR-1 (IRAS 04361+2547) is a deeply embedded Class I protostellar system located in a rotating ring of material in the Taurus molecular cloud (Terebey et al. 1990). Near-infrared imaging and millimeter interferometry show a bipolar outflow extending southeast to the northwest (Terebey et al. 1990). A dark absorption band is observed at 2.2 $\mu$m perpendicular to the direction of the reflection nebulosity, that Terebey et al. interpret as high density circumstellar material seen in absorption against the extended continuum. The spectral energy distribution of TMR-1 peaks at 60$\mu$m and it shows absorption in the 10$\mu$m silicate band, similar to other embedded protostellar systems (Terebey et al. 1990). Terebey et al. estimate an $A_{v}$ of 28$\pm$2 mag, a $T_{eff}$ of 5600$\pm$400 K, a stellar mass of $\sim$0.5$M_{\sun}$ and a bolometric luminosity of $\sim$3$L_{\sun}$ for this source. 

Terebey et al. (1998; hereafter T98) presented {\it HST}/NICMOS observations of TMR-1, and were able to resolve this protostar into two point sources, A and B (the northern component was named A), at a separation of 0.31$\arcsec$. A third fainter point source, TMR-1C, was detected at a separation of about 10$\arcsec$ ($\sim$1000 AU) from the protobinary. Assuming that TMR-1C is located in Taurus and is at the same age as TMR-1AB (0.3 Myr), T98 estimated a bolometric luminosity of approximately 10$^{-3}$$L_{\sun}$ and a mass of 2-5 $M_{J}$ for this source, thus classifying it as a candidate protoplanet. More interestingly, their observations revealed a narrow filament-like structure extending south-east from the central proto-binary system towards TMR-1C. T98 suggested a morphology in which the candidate protoplanet may have been ejected from the TMR-1 system, with the filament traversing the ejection path. Follow-up low-resoluion ({\it R} $\sim$120) near-infrared spectroscopy obtained by Terebey et al. (2000) with the Keck/NIRC instrument did not show any absorption features. In comparison with a grid of extincted M dwarf spectra, they determined a spectral type of M4.5, the latest that could be assigned to TMR-1C. When placed in a theoretical H-R diagram, the object was found to straddle the sub-stellar boundary at masses of $\sim$0.08$M_{\sun}$, with an effective temperature higher than that predicted for giant planets by theoretical models. However, given the low quality of the spectrum and our lack of knowledge on the atmospheres of very young planets, their observations remained inconclusive about the true nature of this object, and there remained the possibility that TMR-1C is a low-luminosity background star. 

We present here two epochs of near-infrared photometric observations for the TMR-1 system. The time span of $\sim$7 years between the two sets of observations provides with an opportunity to check for any photometric variability that would indicate the youth of this source, as well as to determine the proper motion of the components.

\section{Observations \& Data Reduction}

Observations in the {\it H} and $K_{s}$ bands were obtained on 24th October, 2002, at the CFHT 3.6m telescope with the CFHTIR infrared camera (RUNID: H41D; PI: E. L. Mart\'{i}n). This infrared camera has a field of view of $3.6\arcmin\times3.6\arcmin$, with a pixel scale of 0.211$\arcsec$ pixel$^{-1}$. The exposure times were 60s and 20s in the {\it H} and $K_{S}$ bands, respectively. A dither pattern of four positions was requested, where the dither sizes were between 30$\arcsec$ and 60$\arcsec$, and 6 exposures were obtained at each dither position. The raw frames were bias-corrected and flat-fielded using basic IRAF/{\it ccdred} routines. For each object frame, a sky frame was constructed by median-combining all of the dithered observations excluding that object frame, without making any corrections for the offsets. The final sky-subtracted calibrated frames were first aligned to a common $x$ and $y$ position using the IRAF task {\em imshift}, and then averaged using the task {\em imcombine}. Image astrometry was obtained using the IRAF tasks {\it ccfind}, {\it ccmap}, {\it ccsetwcs} and {\it cctran} under the {\em imcoords} package. The errors in astrometry are 0.08$\arcsec$ in RA and 0.06$\arcsec$ in Dec for the {\it H}-band observations, and 0.07$\arcsec$ in RA and 0.08$\arcsec$ in Dec for the $K_{s}$ data. The observing conditions were photometric for this night and the seeing varied between 0.7$\arcsec$ and 0.8$\arcsec$ in both bands. 

We obtained another set of observations in the $K_{s}$ band on 20th January, 2009, at the CFHT 3.6m telescope, using the near-infrared wide-field imager, WIRCam. These observations were obtained under the Director's Discretionary Time (RUNID: 08BD84). WIRCam covers a 20$\arcmin\times20\arcmin$ field of view, with a pixel scale of 0.3$\arcsec$/pix. The night was photometric with a seeing of 0.5$\arcsec$ in $K_{s}$. We requested 5 dither positions, with five exposures of 20s each to be obtained at each dither position. The raw images were pre-processed by the WIRCam observer team using the `I'iwi data reduction pipeline (version 1.9). The `detrended' images obtained from the pipeline first have all detector imprints removed, followed by dark subtraction, flat-fielding and sky-subtraction. The crosstalk is removed from the final pre-processed images, and astrometry is performed with an rms scatter of the resulting WCS solution between 0.3$\arcsec$ and 0.8$\arcsec$. The `l'iwi pipeline processed data were shifted and combined using the IRAF task {\em imshift} and {\em imcombine}. 

Aperture photometry was performed on the combined image for each band using the IRAF task {\em apphot} under the {\em digiphot} package. The data were calibrated using 2MASS magnitudes for bright sources in the same fields. The radial profiles for these bright sources were checked for any saturation effects. Table~\ref{table1} lists the photometry from our 2002 and 2009 observations, along with the photometry reported by T98 from their {\it HST}/NICMOS observations. TMR-1AB was saturated in our WIRCam/2009 observations. We have a $K_{s}$=11.3 mag for this source, while the WIRCam saturation limit is 13.8 mag. It is also close to saturation in the 2002 $K_{s}$-band observations. The radial profile for this protobinary does not show any saturation effects in the 2002 {\it H}-band data. We note that we could not resolve TMR-1AB into its individual components, and the composite 2002 {\it H}-band photometry is listed in Table~\ref{table1}. The photometry for TMR-1C is not affected by saturation in any of the observations.  

Persson et al. (1998) developed a grid of {\it J-, H-, K-}, and $K_{s}$-band standards for the {\it HST} NICMOS camera, using observations from the Las Campanas Observatory (LCO) in Chile. Later, Carpenter (2001) derived 2MASS-LCO transformation equations using 2MASS photometry for 82 stars from Persson et al. (1998). Equations (1) and (2) below are the color transformation relations from Persson et al. (1998) and Carpenter (2001), respectively. 

\begin{equation}
(H-K)_{CIT} = (0.974 \pm 0.020)(H-K)_{LCO} \pm (0.013) 
\end{equation}

\begin{equation}
(H-K_{s})_{2MASS} = (1.019 \pm 0.010)(H-K)_{LCO} + (0.005 \pm 0.005)
\end{equation}

In Table \ref{table1}, we have listed our CFHTIR/2002 2MASS-calibrated ($H-K_{s}$) color transformed to $(H-K_{s})_{LCO}$, using Eqn. (1). The results are similar using Eqn. (2). We find a difference in the $(H-K_{s})$ color of $\sim$0.01 mag for TMR-1C, and 0.02-0.03 mag for TMR-1AB.

\section{Analysis}
\subsection{Astrometry}
\label{astrometry}

Table~\ref{table2} lists the positions for TMR-1AB and TMR-1C from our 2002 and 2009 observations, and the {\it HST} positions from T98. The separation between TMR-1AB and TMR-1C is 9.8$\arcsec$ in the 2002 observation, which is consistent with the 10$\arcsec$ separation reported by T98. The position uncertainty is 0.08$\arcsec$ in RA and 0.06$\arcsec$ for the CFHTIR/2002 images. The error in astrometry as reported from the WIRCam pipeline (for the 2009 data) is between 0.3$\arcsec$ and 0.8$\arcsec$. The 1-$\sigma$ position error reported by T98 is 0.35$\arcsec$. 

\begin{figure*}
\centering
 \includegraphics[width=7cm]{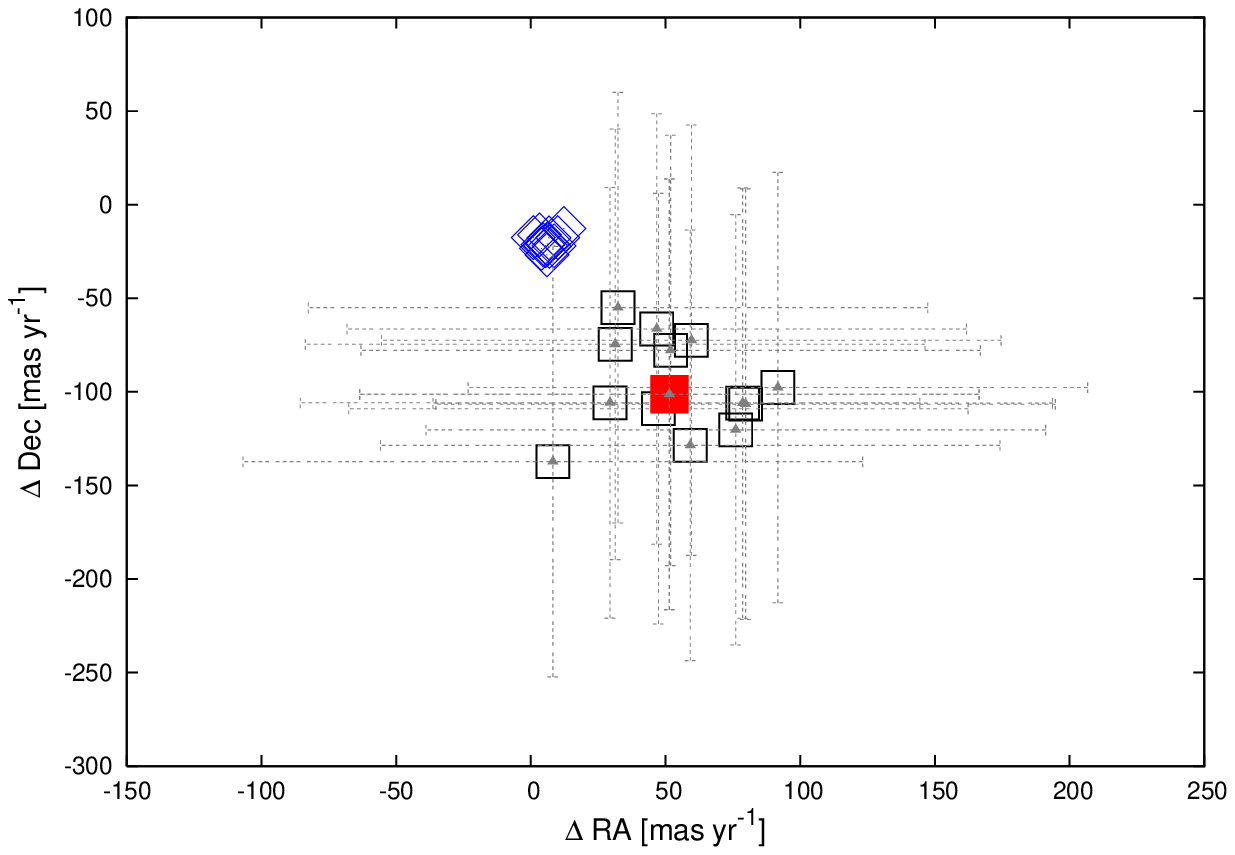}  \\
 \includegraphics[width=7cm]{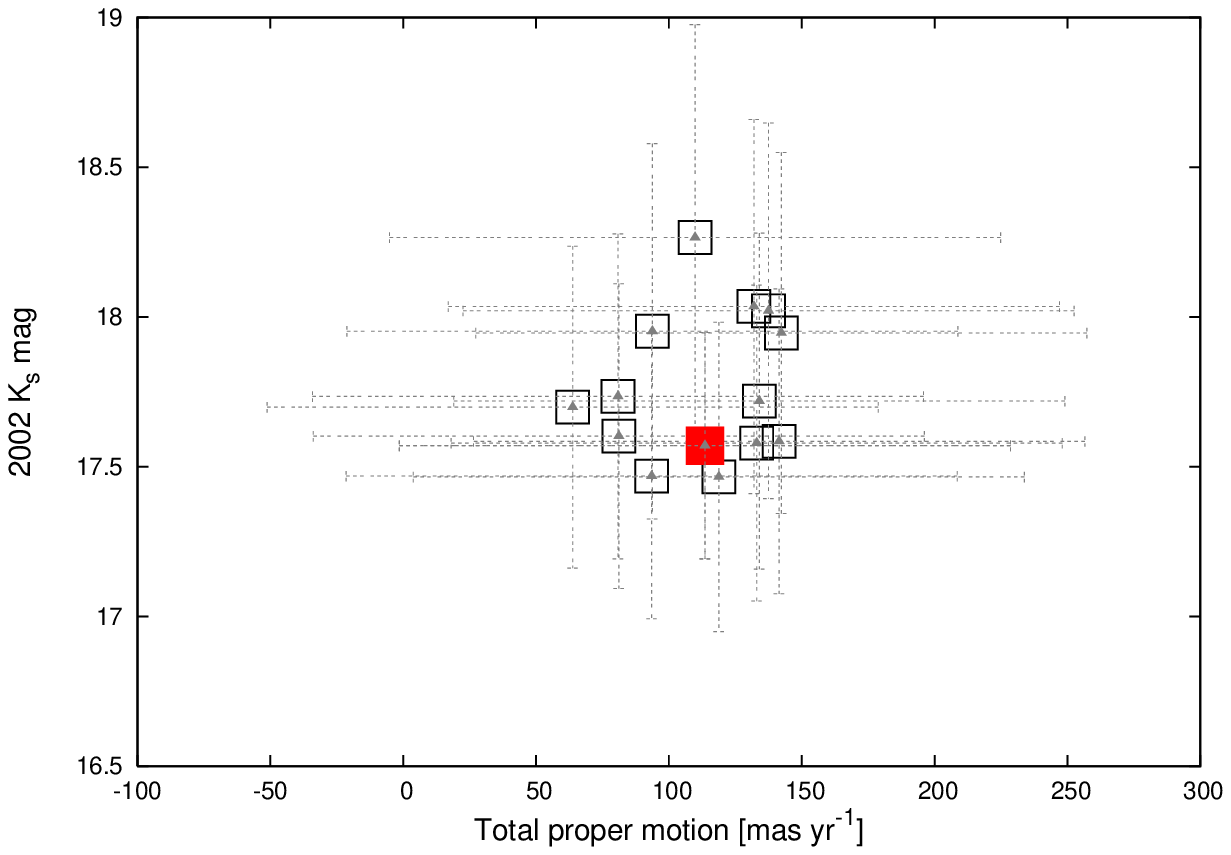} \\ 
 \caption{ {\it Top}: Proper motion in RA vs. proper motion in Dec. Proper motions have been obtained from the 2002 and 2009 $K_{s}$-band positions. Red square is TMR-1C. Grey squares are faint objects with $K_{s} <$ 17 mag, that are common in the 2002 and 2009 $K_{s}$-band images. The uncertainties are large due to the 0.3$\arcsec$-0.8$\arcsec$ position uncertainty for the 2009 observations. Blue squares are the median proper motions for the 11 groups in Taurus from Luhman \& Mamajek (2010). The errors on proper motions for these groups are between 1 and 2 mas/yr. The typical proper motion for Taurus from Ducourant et al. (2005) is +2,-22 mas/yr, with uncertainties of 2-5mas/yr. {\it Bottom}: The 2002 K$_{s}$ photometry vs. the total proper motion. The total proper motion for the 11 groups in Taurus from Luhman \& Mamajek (2010) is between 16 and 27 mas/yr. From Ducourant et al. (2005), the typical proper motion in Taurus is 22.1 mas/yr.  }
  \label{fig1}
\end{figure*}

Figure \ref{fig1} plots the proper motion for TMR-1C, as determined from the 2002 and 2009 $K_{s}$ observations (using a 7 year time span). We have also plotted the proper motions for other faint sources that are common in these two $K_{s}$-band images, and that have $K_{s} <$ 17 mag. The large uncertainty on the proper motion is due to the 0.8$\arcsec$ position error for the 2009 observations. Also included for comparison are the median proper motions for the 11 groups in the Taurus-Auriga star-forming region, discussed in Luhman \& Mamajek (2010). The errors on proper motions for these groups are between 1 and 2 mas/yr. The typical proper motion for Taurus from Ducourant et al. (2005) is +2,-22 mas/yr, with uncertainties of 2-5 mas/yr. 

Due to the saturation of TMR-1AB in the 2009 observations, it is difficult to determine its proper motion using these two sets of observations. We have calculated the proper motion for TMR-1AB from the 2MASS and the 2002 {\it H}-band positions. The proper motion for TMR-1AB is found to be 90$\pm$32 mas/yr (using a 4 year time span). The astrometry and the proper motion for TMR-1AB however is unreliable, considering that it is for an unresolved binary system, and a baseline of 4 years is not large enough to obtain a good estimate on the proper motion. 

From Fig.~\ref{fig1}, the proper motion for TMR-1C seems consistent with the Taurus locus, within the large error bars. The astrometric results however are not conclusive given the large uncertainties. The proper motions for TMR-1AB and TMR-1C are a factor of $\sim$5 larger than the expected proper motions for the Taurus members. Further observations are required to confirm the common proper motion of the components, as well as to confirm their Taurus membership. A longer baseline is needed for useful proper motion determination. We can estimate that about 20 more years are needed to significantly improve our results. 

While accurate estimates on the proper motion of the components cannot be obtained with the observations at hand, another observational test to distinguish whether TMR-1C is a young protoplanet or an old background star is the confirmation of photometric variability, similar to YSO variability, that could be associated with the presence of circumstellar material. We discuss this further in the next section.

\section{Discussion}
\subsection{Photometric Variability for TMR-1C: Blue ($H-K_{s}$) Colors}

\begin{figure*}
\centering
 \includegraphics[width=7cm]{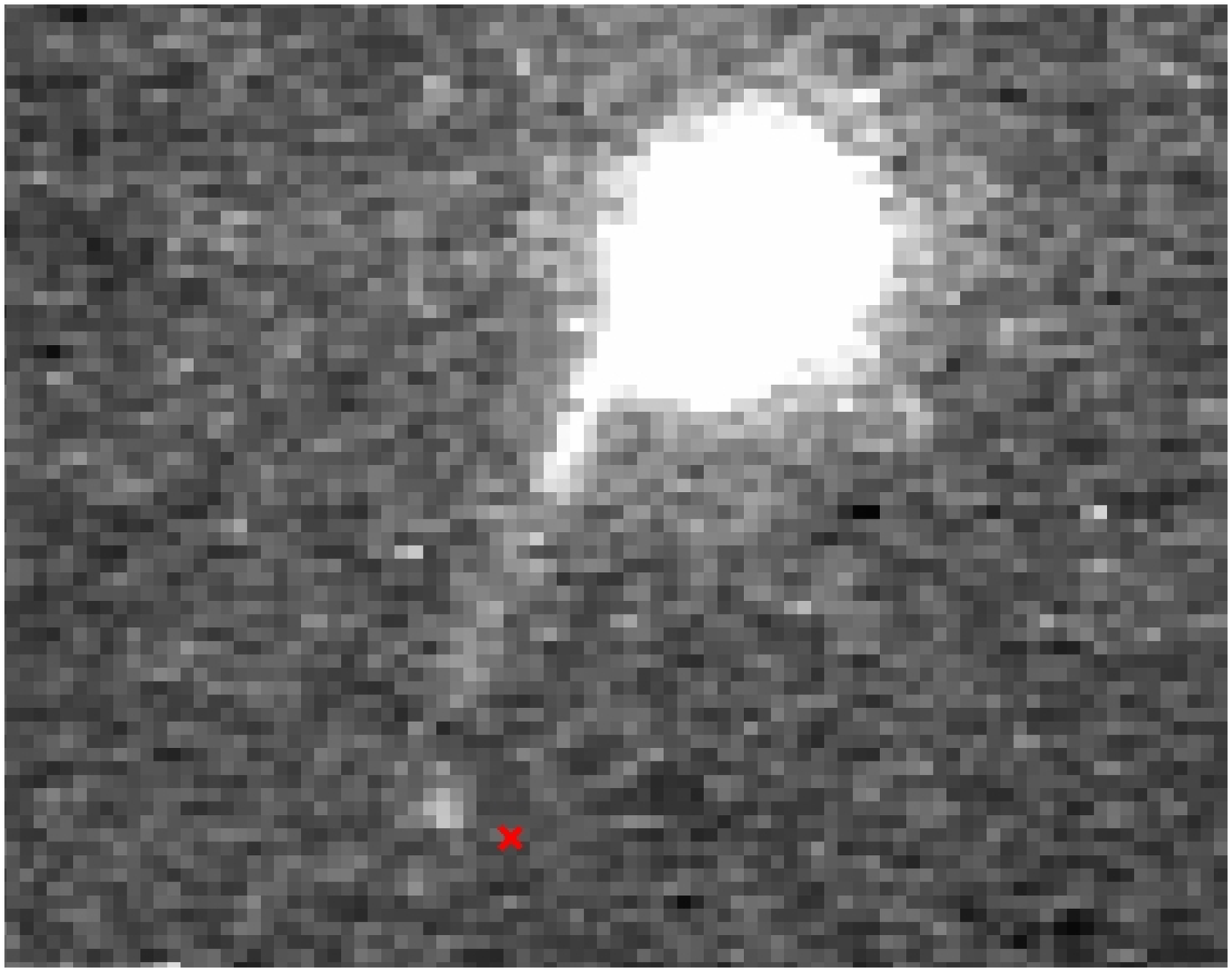}  \hspace{0.2cm}
 \includegraphics[width=7cm]{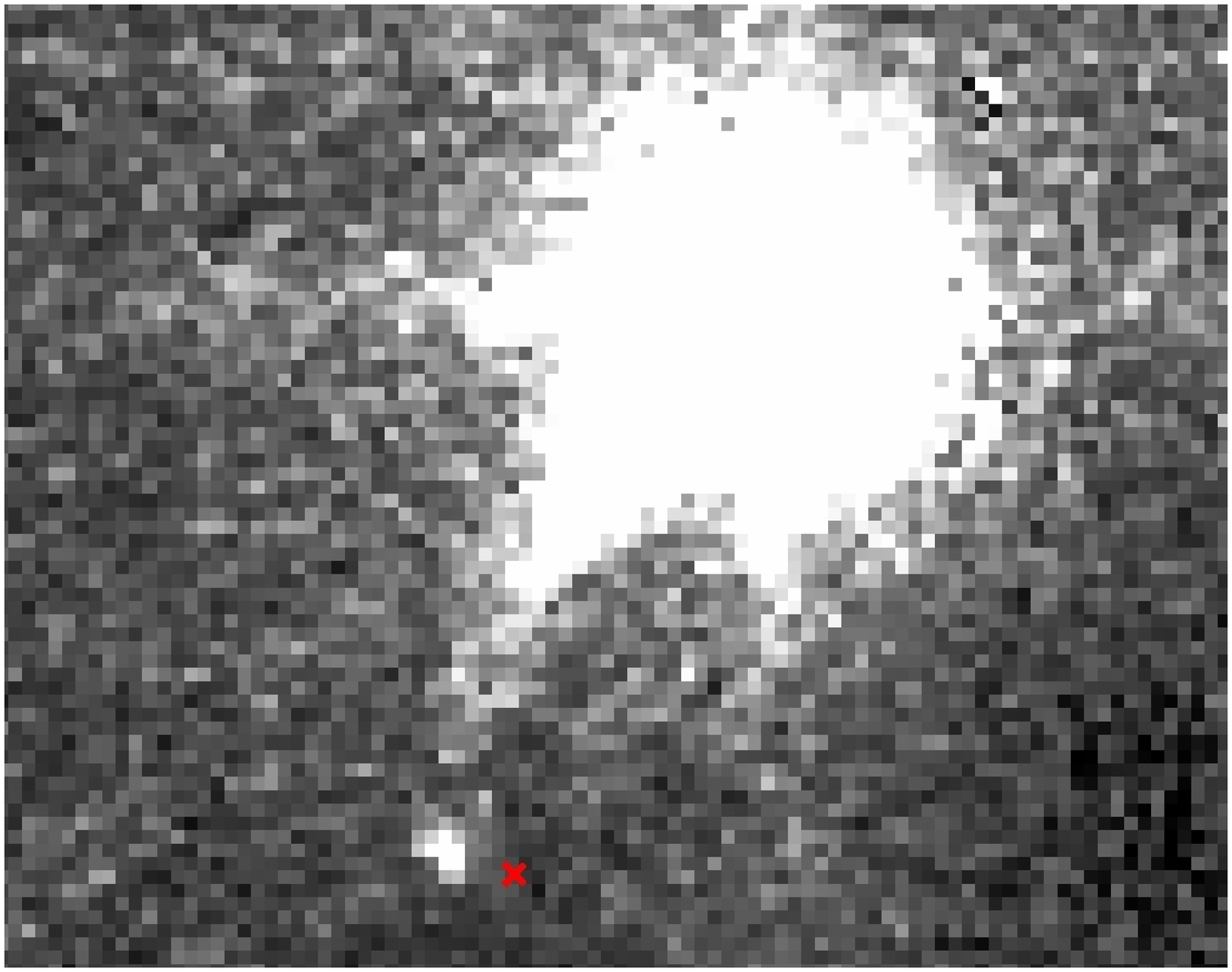} \\  \vspace{0.2cm}
  \includegraphics[width=6.5cm]{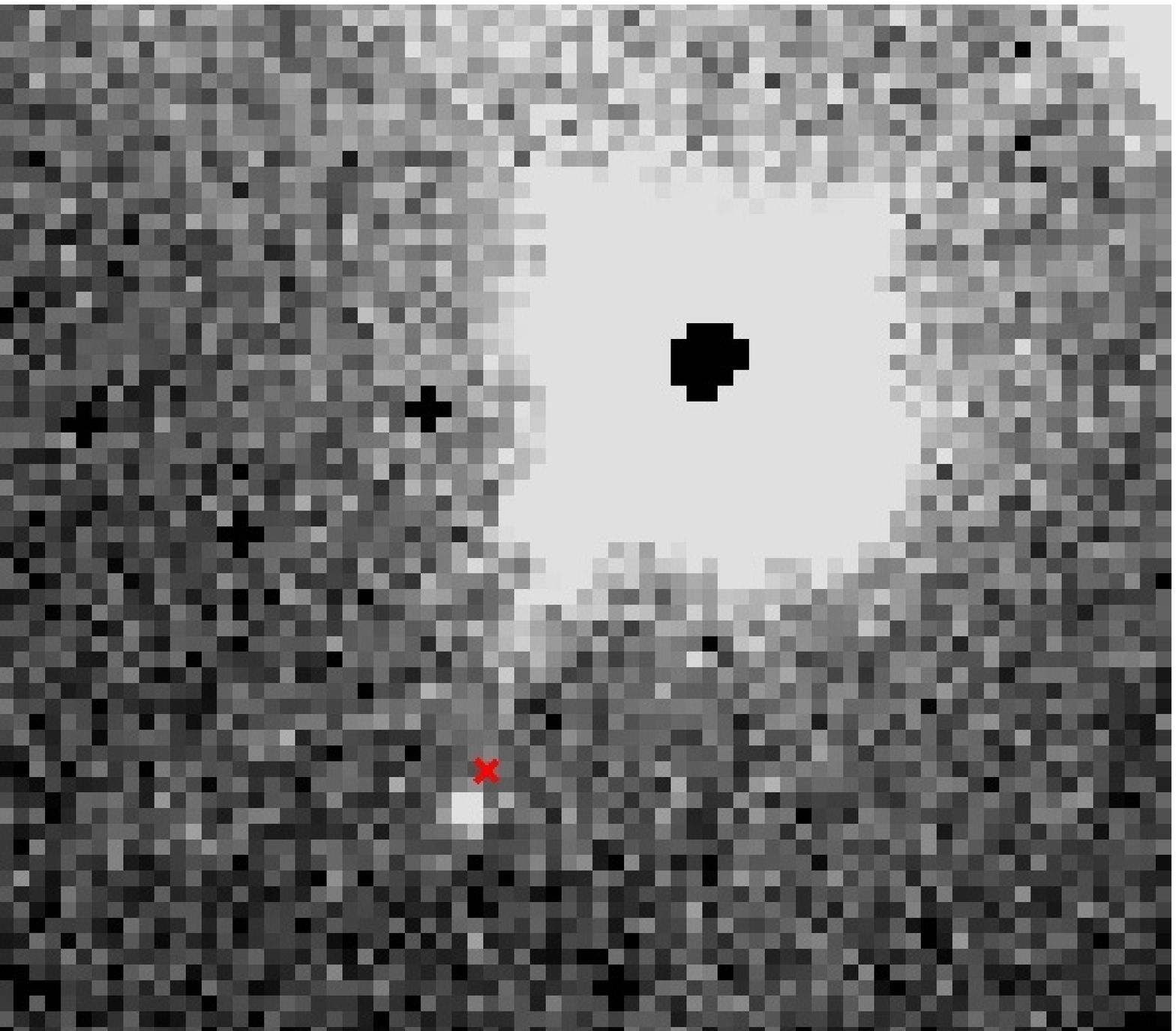}
 \caption{Top panel shows the CFHTIR/2002 {\it H}-band ({\it left}) and $K_{s}$-band ({\it right}) image. Red cross marks the {\it HST}/NICMOS position for TMR-1C from T98. Both images are 14$\arcsec$ on a side. Bottom panel shows the WIRCam/2009 $K_{s}$ image. Here, the red cross marks the CFHTIR/2002 position for TMR-1C. This image is 20$\arcsec$ on a side. In all three images, north is up, east is to the left.}
  \label{fig2}
\end{figure*}

TMR-1C was clearly detected in both of our {\it H}- and $K_{s}$-band observations (Fig. \ref{fig2}). A comparison of the F160W (1.6$\mu$m) {\it HST} photometry from T98, and the {\it H}-band CFHT/2002 photometry indicates strong variability of 1.53$\pm$0.5 mag. In the $K_{s}$-band, the variability is 0.8$\pm$0.3 mag when comparing the F205W photometry from T98 and our 2009/WIRCam observations. We thus find strong evidence of variability at a level of $>$2-$\sigma$ in both the {\it H} and $K_{s}$-bands. 

From our 2002 observations, we find a ($H-K_{s}$) color of 0.3 mag for TMR-1C. This is a much bluer color than the value of 1.3 mag reported by T98 from {\it HST} observations. The ($H-K_{s}$) color thus shows strong variability of $\sim$1 mag over this $\sim$4 yr time span. Also, we observe brightening in both the {\it H}- and $K_{s}$-bands when the colors are bluer, i.e. the object gets redder as it becomes fainter. Since we have only these two measurements for the ($H-K_{s}$) color, we cannot accurately determine the variability timescale, and it may be much shorter than the 4-year period considered here. 

\subsection{Possible Origins for the Variability}
\subsubsection{Starspots}

To explain the observed variability, we look into the origins of photometric variability in young stellar objects, as discussed in Carpenter et al. (2001; C01). Starspots, whether hot or cold, cause objects to become redder as they get fainter. We can reject the presence of cool spots for TMR-1C, due to the low variability amplitudes. Cool spots are produced due to the interaction of the magnetic field with the photospheric gas, and are thus an indicator of magnetic activity (e.g., Scholz et al. 2009). C01 estimate a maximum change in the ($H-K_{s}$) color of $\sim$0.03 mag for cool spots, a factor of 10 smaller than that observed for TMR-1C. Hot spots can produce larger variability amplitudes. These are the regions on the stellar surface where accreting material from the disk falling along the magnetic lines of force impacts onto the star. Hot spots are thus related to the accretion activity, and any variations in the mass accretion rate or the accretion flow geometry can result in irregular variability, that can last for as short as a few days or weeks (C01). We have used the relation provided in C01 [Eqn. (7)] to calculate the photometric amplitudes expected from hot spots. Typical values for filling factors among young stellar objects are $<$30\%, with a temperature contrast of 10-30\% between spot and photosphere (e.g., Bouvier et al. 1993). For very low-mass stars with $M<$0.4$M_{\sun}$, Scholz et al. (2005) have reported temperature contrast of 18-31\% with low filling factors of 4-5\%. We set the effective temperature to 1700 K for TMR-1C, and calculated the relative variability, $\Delta$$(H-K_{s})$/$\Delta H$, for a range of filling factors between 1\% and 50\%, and spot temperatures between 2000K and 10,000K. The relative variability thus calculated ranges between $\sim$0.18 to 0.47. For TMR-1C, $\Delta$$(H-K_{s})$/$\Delta H$=0.66. Even if we reduce the filling factor to 0.1\%, we cannot obtain such large relative color changes. Thus hot spot activity cannot explain the observed variability for TMR-1C, since the amplitudes of variations are too large.

\subsubsection{Disk Emission}

Another possible origin is variability in the disk emission. The presence of warm circumstellar dust can produce strong excess emission in the {\it H}- and $K_{s}$-bands (e.g., Doppmann et al. 2005). The maximum temperature that the dust can reach depends very strongly on the local gas density, and temperatures as high as 1500 K could be reached if the gas density is high (e.g., Kama et al. 2009). Accretion-induced veiling from the disk onto the protoplanet could produce a NIR excess emission (e.g., Hartigan et al. 1995). Many young sub-stellar/planetary mass objects are found to posses active accretion disks (e.g., Barrado y Navacu\'{e}s \& Mart\'{i}n 2003; Luhman et al. 2005), making it likely for TMR-1C to be a disk candidate. A variable accretion rate could produce variability in the observed NIR colors (e.g., Beck 2007), or changes in the inner disk structure, such as an inner hole, could reduce the dust emission. However, disk emission will be stronger in the $K_{s}$- band than in the {\it J}- or the {\it H}-bands, making the system bluer as it gets fainter (C01). Comparing the 2002 and 1998 photometry, TMR-1C shows larger variability in the {\it H}-band than the $K_{s}$- band; it has brightened by 1.5 mag in the {\it H}, compared to 0.5 mag in the $K_{s}$. Thus disk emission also cannot be a possible explanation for the observed variability.

\subsubsection{Variable Extinction}
\label{varext}

The most likely scenario for TMR-1C seems to be variable extinction, due to inhomogeneities in the circumstellar environment or the ambient molecular cloud that is intercepted by the line of sight. Variable extinction can cause larger color changes than hot spots, and would make the object redder as it gets fainter (C01). For TMR-1AB, Terebey et al. (1990) estimated an extinction from 2.5 to 4 at {\it K} ($A_{V}$$\sim$22-36 mag). T98 suggested that the extinction is likely less towards C, at a 10$\arcsec$ distance from the protostar. The total {\it H}-band amplitude for TMR-1C (from 1998 and 2002 observations) corresponds to $\Delta A_{V}$=8.7 mag. The total $K_{s}$ amplitude corresponds to $\Delta A_{V}$=4.6 mag. The change in extinction is thus between $\sim$5 and 9 mag over a $\sim$4 year period. C01 have discussed the possible inhomogeneities in the circumstellar disk that can occult the star to cause extinction variations. The inhomogeneities could be due to e.g., azimuthal asymmetries in the plane of the disk, or a flared outer disk. Azimuthal asymmetries require an inclination of the disk close to edge-on. A face-on inclination is unlikely to cause color variations due to varying extinction. An $A_{V}$$\sim$22-36 mag indicates considerable extinction towards this system in the line of sight. One way to distinguish between extinction variations caused by the ambient molecular cloud and the circumstellar disk is the longer timescale expected for cloud transit events (C01). With only two epochs of data, we cannot estimate the variation timescale accurately. Nevertheless, extinction by an edge-on disk or an outer flared disk, or the surrounding molecular cloud is a plausible explanation for the large amplitudes of variations observed in TMR-1C.

In the scenarios discussed above, we have considered TMR-1C to be a young disk source, surrounded by an actively accreting disk and possibly even an infalling envelope. The red and flattish spectrum obtained by Terebey et al. (2000) and the photometric variability from our observations indicate that TMR-1C has a non-photospheric contribution to the continuum. Based on its ejection velocity and the current separation from TMR-1AB, T98 estimated that TMR-1C must have been recently ejected, about 1000 years ago. So this is a young protoplanet, and being ejected from a Class I protostellar system, is likely to be surrounded by optically thick envelope and disk material. Doppmann et al. (2005) estimate an intrinsic color of ($H-K_{s}$) = 0.6 $\pm$ 0.4 for Class I sources. They however consider this estimate to be highly uncertain considering the difficulty in disentangling the contribution from the surrounding envelope and disk material and the scattered emission. It may be that we have measured the intrinsic colors for TMR-1C in our measurement, while the T98 photometry is more affected by extinction. 

There may also be the possibility that this object is not a young disk source, but some other type of variable object. Certain background objects could also be variable (e.g., Minniti et al. 2010), or quasars are also known to be optically variable, with variation amplitudes being larger than $\sim$0.5 mag in some cases (e.g., Hawkins \& Veron 1993). 

\subsubsection{A system similar to KH 15D?}
\label{kh15d}

It is interesting to compare TMR-1C with the pre-main sequence object KH 15D, that shows blueing of the ({\it J-H}) and ($H-K_{s}$) colors during eclipsing over a $\sim$48 day period, while outside of the eclipse, redder colors are observed (Kusakabe et al. 2005). A model presented by Winn et al. (2006) suggests that KH 15D is an eccentric binary system being occulted periodically by the edge of a precessing circumbinary disk. During an eclipse, the photosphere of the star is totally occulted, and the blueing observed is due to emission from a halo around the star. This halo could be scattered starlight from small particles in the upper layers of the circumbinary disk. If such a scenario exists for TMR-1C, it suggests that our measurement was made during the eclipsing period that resulted in a bluer color, while T98 observation of a redder color was made outside of the eclipse. However, there is no evidence that TMR-1C is a binary itself. Also, it is too much to invoke a common disk around TMR-1ABC, since the distance is too large ($\sim$1000AU). It is more likely to have a circumbinary disk around TMR-1AB and a circum(sub)stellar disk around TMR-1C. If the disk around TMR-1C is highly inclined ($\S$\ref{varext}) then this could cause variability in colors. One caveat to this is that KH 15D shows bluer colors when it is fainter, contrary to TMR-1C. KH 15D is a K7 star, more massive than TMR-1C. In the case of KH 15D, the object becomes blue when it is fainter because the central star is occulted by the disk, and the scattered light is bluer than the central star. But we can have the contrary in the case of TMR-1C because the scattered light is redder than the central object, which itself may have blue near-infrared colors due to the presence of methane in the photosphere. There are thus some possible analogies between this system and KH 15D.

\subsection{T dwarf Colors}

\begin{figure*}
\centering
 \includegraphics[width=14cm]{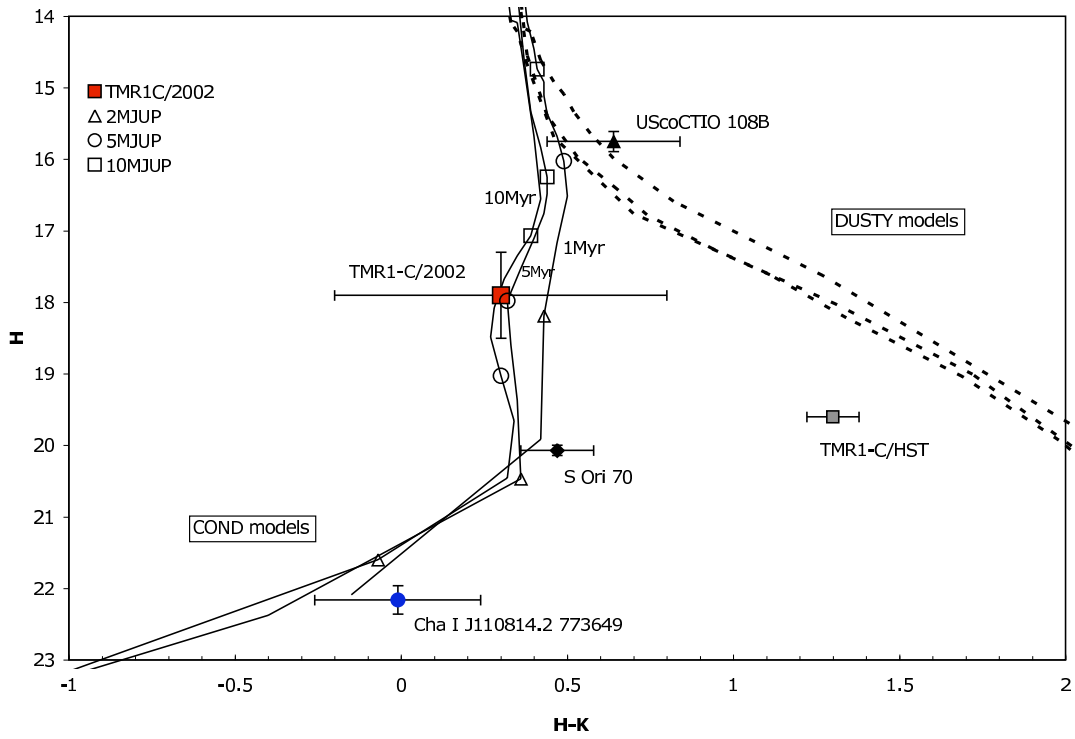} \\ 
 \caption{The near-infrared color-magnitude diagram for TMR-1C. Also included are three very-low mass/plnateray mass objects. The solid and dashed lines are COND and DUSTY models, respectively, from Allard et al. (2001), for ages of 1, 5 and 10 Myr.  }
  \label{fig3}
\end{figure*}

Figure~\ref{fig3} shows the ($H-K_{s}$) vs. $K_{s}$ color-magnitude diagram (cmd) for TMR-1C. Overplotted are the DUSTY and condensed atmosphere (COND) models from Allard et al. (2001). Included for comparison are three very low-mass/planetary mass objects located in nearby young clusters. S Ori 70 is a well-known mid-T dwarf with an estimated mass of 2-5 $M_{Jup}$, and has been confirmed as a member of $\sigma$ Ori cluster ($\sim$3myr; Zapatero Osorio et al. 2002) based on its proper motion and near- and mid-infrared colors (Zapatero Osorio et al. 2008; Mart\'{i}n \& Zapatero Osorio 2003). UScoCTIO108B is a $\sim$14 $M_{Jup}$ mass object in the $\sim$5 Myr old Upper Scorpius association, and forms a wide ($\sim$670 AU) binary pair with UScoCTIO108A (B\'{e}jar et al. 2008). Cha I J110814.2-773649 is a faint object discovered by Comer\'{o}n et al. (2004) in the direction of the Cha I star-froming region, and shows a very blue ($H-K_{s}$) color of -0.01. Comer\'{o}n et al. estimate a mass of 0.7 $M_{Jup}$ for this source if it is at an age of 2 Myr, or $\sim$1$M_{Jup}$ if it is a 5 Myr old object. The models by Allard et al. (2001) are two limiting cases of dust treatment in the atmosphere. The DUSTY models include grain formation as well as grain opacities. The resulting spectral distribution thus appears redder with weaker molecular opacities, as compared to the models that do not consider grain opacities. The COND models do consider grain formation but do not take into account the grain opacities, resulting in a spectral distribution that is transparent blueward of 1.0$\mu$m, and very similar to the dust-free case. From the cmd shown in Fig.~\ref{fig3}, the blue ($H-K_{s}$) color of 0.3 for TMR-1C could be explained by the presence of a condensed atmosphere, as observed among T dwarfs. The location of TMR-1C indicates a mass of  $\sim$5 $M_{Jup}$. Though the uncertainties are large, this is consistent with the 2-5 $M_{Jup}$ estimate made by T98. The location of the redder color of ($H-K_{s}$)=1.3 mag from T98 photometry indicates an earlier mid-L type, as it lies close to the blue-turnover point observed among mid-L to early-T dwarfs (e.g., Marley et al. 2002), and thus a higher mass for this source. 

\subsection{Young Wide Very Low Mass Systems}

\begin{figure*}
\centering
 \includegraphics[width=9cm]{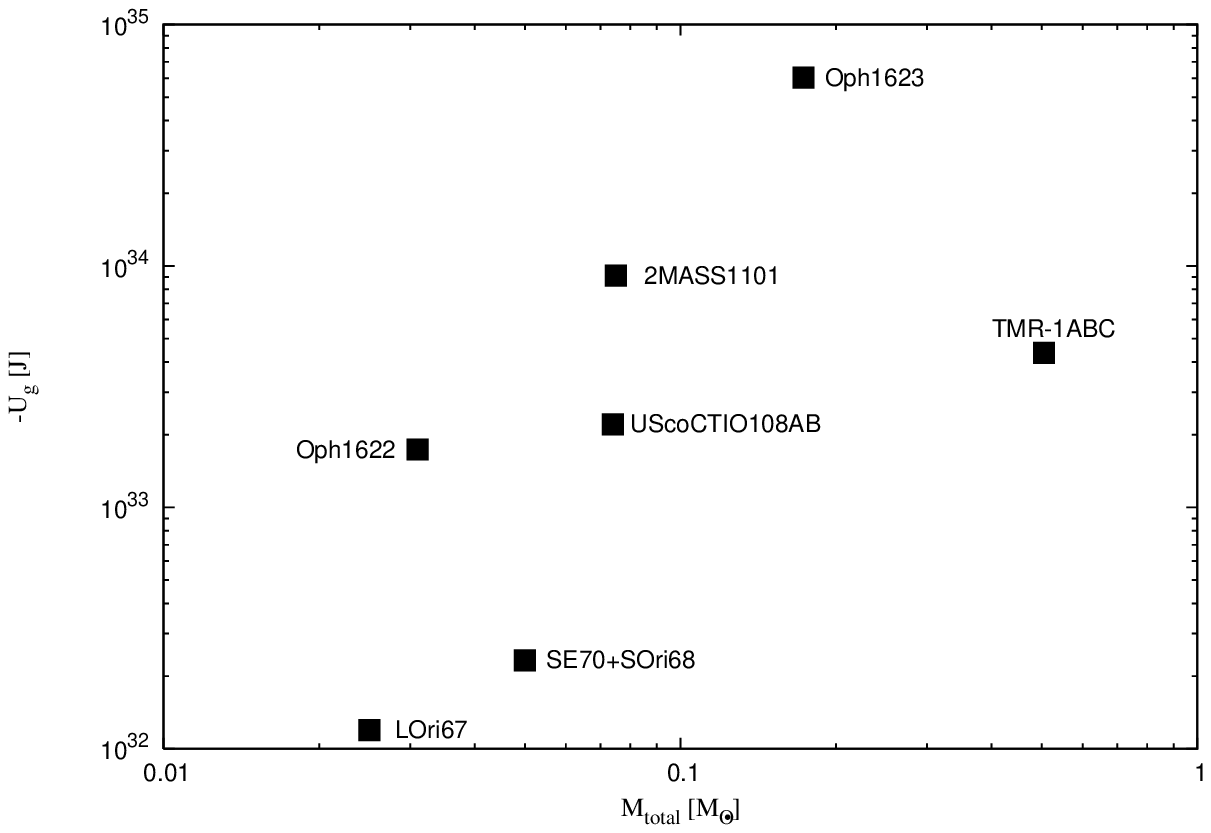} \\  \vspace{0.2cm}
 \includegraphics[width=9cm]{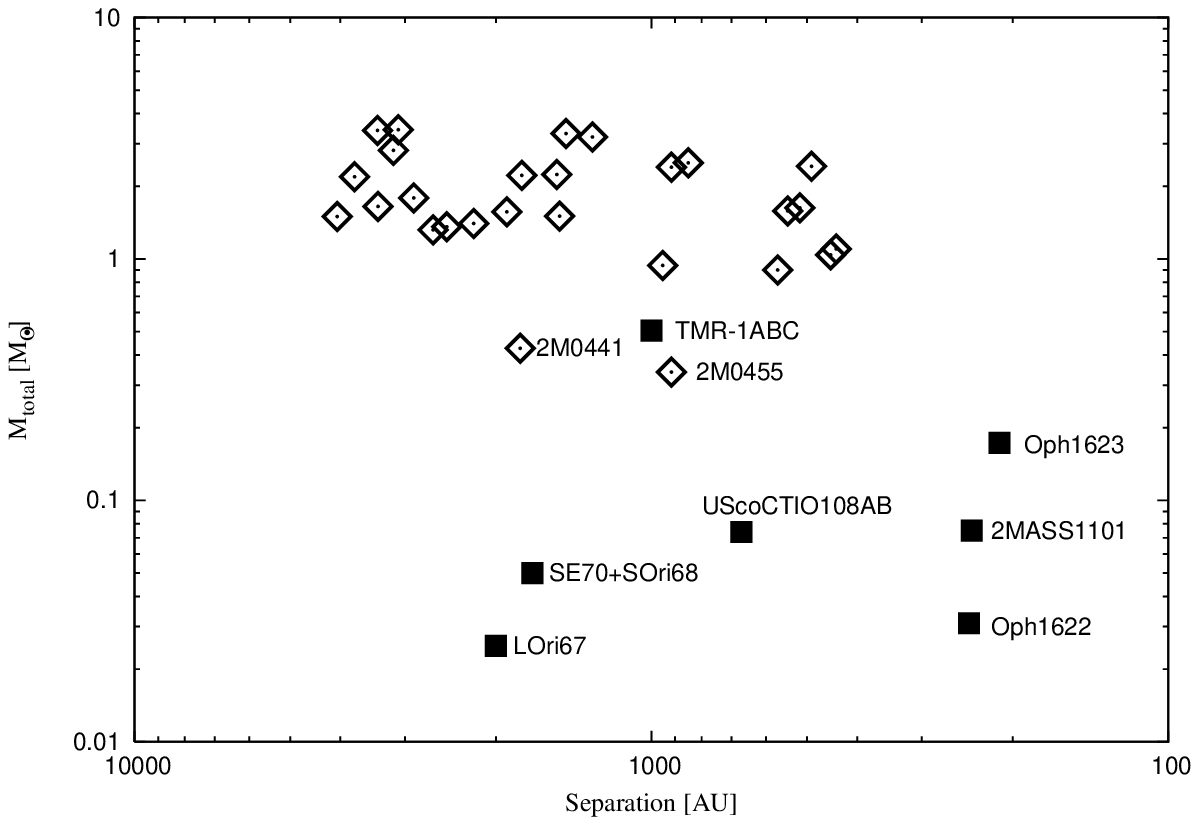} \\  
 \caption{{\it Top}: (a) The binding energy for TMR-1ABC compared with other wide systems in nearby young clusters with a sub-stellar/planetary-mass secondary. {\it Bottom}: (b) Total system mass as a function of separation in AU for TMR-1ABC and other low-binding energy systems (bold squares; see text for details). Included for comparison are wide (500-5000 AU) systems in Taurus (plotted as diamonds) from Kraus \& Hillenbrand (2009). }
  \label{fig4}
\end{figure*}

A few other wide binaries containing a planetary mass component have been discovered in young nearby clusters, such as the pair of SE 70 ($\sim$45 $M_{Jup}$) and S Ori 68 ($\sim$5 $M_{Jup}$) at a projected separation of 1700 AU in the $\sigma$ Orionis cluster (Caballero et al. 2006), the wide pair UScoCTIO 108AB ($\sim$60 and $\sim$14 $M_{Jup}$) at a projected separation of $\sim$670 AU in Upper Scorpius (B\'{e}jar et al. 2008), Oph 16222-2405, the 243 AU wide pair of $\sim$17 and $\sim$14 $M_{Jup}$ components in Ophiuchus (Close et al. 2007; Luhman et al. 2007), and the LOri67 system in Lambda Orionis made up of a $\sim$17$M_{Jup}$ primary and an $\sim$8$M_{Jup}$ secondary at $\sim$2000 AU projected separation (Barrado y Navacu\'{e}s et al. 2007). In Figure~\ref{fig4}a, we have compared the gravitational potential energy (binding energy), $U_{g} = - G M_{1} M_{2} / r$, for TMR-1ABC with these young wide systems, using {\it r} as the projected separation. Also included are the Oph1623-2402 system ({\it r}= 212 AU; Close et al. 2007) in Ophiuchus and the 2MASS J1101192-773238 system ({\it r}= 240 AU; Luhman 2004) in Chamaeleon I, that have relatively higher mass secondaries (sub-stellar) but the total mass of the system is still very small ($\leq$ 0.1$M_{\sun}$). These systems are among the lowest binding energy systems for their total masses in the solar neighborhood (e.g., Caballero 2009). It is estimated that approximately 6\% of such young ($<$ 10 Myr) very low-mass binaries (VLM; $M_{total} < 0.2 M_{\sun}$) are at wide separations of $>$100 AU, and this frequency is a factor of $\sim$20 higher than that found among older wide VLM systems in the field (e.g., Close et al. 2007, and references therein). As these authors explain, VLM binaries dynamically dissipate or evaporate over time due to e.g., stellar encounters. However, the probability of such encounters will be lower if the stellar density is low. The frequency of wide (500-5000 AU) systems at a given mass is found to be higher in the less dense regions like Taurus and Chamaeleon I, compared to the relatively denser Upper Scorpius association (Kraus \& Hillenbrand 2009), which suggests a higher chance of survival for such low binding energy systems in less dense regions. However, there is a paucity of wide VLM systems in Taurus. Figure~\ref{fig4}b plots the total system mass as a function of the projected separation. We have included in Fig.~\ref{fig4}b wide (500-5000 AU) systems in Taurus from the work of Kraus \& Hillenbrand (2009; and references therein). The known young wide VLM systems mentioned above are all in the Chamaeleon I, Ophiuchus, Upper Scorpius, or the Orion OB1 association. Most of the Taurus wide systems, however, are among the higher mass members in this region, and there are only two known systems, 2M04554757+3028077 and 2M04414565+2301580, that have $M_{total} \leq 0.3 M_{\sun}$. Kraus \& Hillenbrand (2009) estimate that only $\sim$6\% of all systems (single or binary) in Taurus with total mass between 0.25 and 0.5 $M_{\sun}$ have a wide component in the 500-5000 AU separation range. TMR-1ABC is thus another example of these young ``unusually'' wide systems in Taurus at $M_{total} \leq 0.5 M_{\sun}$. The less dense environment in this region could allow such weakly-bound systems to survive disruption at an early stage.

\section{Summary}
Our observations clearly confirm photometric variability for TMR-1C, similar to that observed among young stellar objects, and provide a strong argument against this faint source being an old background star. The red and flattish spectrum obtained by Terebey et al. (2000) and the photometric variability from our observations indicate that TMR-1C has a significant non-photospheric contribution to the continuum. Building up the SED for this source from optical through mid- and far-IR wavelengths would help to better characterize it, as well as understand the dust extinction and other properties of the occulting material. A longer baseline is needed for useful proper motion determination.

\begin{acknowledgements}

This work was supported by the FP6 CONSTELLATION Marie Curie RTN which is governed by
contract number MRTN-CT-2006-035890 with the European Commission. Based on observations obtained at the Canada-France-Hawaii Telescope (CFHT) which is operated by the National Research Council of Canada, the Institut National des Sciences de l'Univers of the Centre National de la Recherche Scientifique of France, and the University of Hawaii. 

\end{acknowledgements}

\onecolumn
\begin{table}[h]
\centering
\begin{minipage}{20cm}
\caption{Photometry}
\begin{tabular}{ccccccccc}
\hline\hline
 TMR-1       & Observation (Epoch)   & Band/Filter  & Photometry & ($H-K_{s}$)  \\
component &                             &            &  [mag]  & [mag]     \\
\hline

C    & CFHTIR (2002)    &  {\it H}  &  17.9$\pm$0.5& 0.308$\pm$0.6\footnote{Using the color transformation relation from Persson et al. (1998); Eqn. (1)} &  \\
        &  CFHTIR (2002)   &  $K_{s}$ & 17.6$\pm$0.4 & \\
        &  WIRCam (2009)   &  $K_{s}$ & 17.3$\pm$0.3 & \\ \hline
C & {\it HST}/NICMOS (1998)  & F160W (1.6$\mu$m) & 19.43$^{+0.04}_{-0.05}$ &1.31$\pm$0.07&  \\
                                 &   {\it HST}/NICMOS (1998) & F205W (2.05$\mu$m) & 18.12$\pm$0.06 &\\   \hline

AB & CFHTIR (2002)   &  {\it H}  & 11.2$\pm$0.3 &    \\  \hline
A & {\it HST}/NICMOS (1998) & F160W (1.6$\mu$m) & 15.65$^{+0.8}_{-0.5}$ &2.22$\pm$0.7 & \\
                                 &  {\it HST}/NICMOS (1998) & F205W (2.05$\mu$m) & 13.43$^{+0.4}_{-0.3}$ & \\ \hline
B & {\it HST}/NICMOS (1998)   & F160W (1.6$\mu$m) & 17.22$\pm$1.1&3.44$\pm$1.0 &  \\
                                 &  {\it HST}/NICMOS (1998) & F205W (2.05$\mu$m) & 13.78$^{+1.1}_{-0.4}$&  \\  \hline
AB & 2MASS (1998) & {\it H} & 13.02$\pm$0.03 & 2.24$\pm$0.03$^{a}$ \\
       &  2MASS (1998)  & $K_{s}$ & 10.72$\pm$0.02 & \\

\hline
\end{tabular}
\label{table1}
\end{minipage}
\end{table}

 \clearpage
 
\onecolumn
\begin{table}[h]
\centering
\begin{minipage}{20cm}
\caption{Astrometry}
\begin{tabular}{ccccccccc}
\hline\hline
 TMR-1       & Observation  & \multicolumn{2}{c}{Position}    & Sep\footnote{Separation[$\arcsec$] between TMR-1AB and TMR-1C. } & Position Error &  $\mu_{\alpha}$ & $\mu_{\delta}$ & $\mu$\footnote{Proper motion obtained from the 2002 and 2009 $K_{s}$-band positions. Errors on proper motions are 115 mas/yr.} \\
component &                        &     RA (J2000) & Dec (J2000) &                      [$\arcsec$] & [$\arcsec$] & [mas/yr] & [mas/yr] & [mas/yr] \\
\hline

C    & CFHTIR/2002  & 04 39 14.23 & 25 53 12.2    & 9.8 & 0.1 & 51 & -101 & 113  \\
        &  WIRCam/2009   & 04 39 14.22 & 25 53 11.33   &    & 0.3-0.8 & && \\ \hline
C & {\it HST}/NICMOS  & 04 39 14.14 & 25 53 11.8 &  10 & 0.35 & & & \\   \hline

AB & CFHTIR/2002  & 04 39 13.87 & 25 53 21.05  & & 0.1  & & &  \\ \hline
AB & 2MASS  & 04 39 13.89 & 25 53 20.88 & & 0.08 & & & \\ \hline
A & {\it HST}/NICMOS  & 04 39 13.84 & 25 53 20.6 & & 0.35 &\\ \hline
B & {\it HST}/NICMOS  & 04 39 13.83 & 25 53 20.4  & & 0.35& \\

\hline
\end{tabular}
\label{table2}
\end{minipage}
\end{table}

\clearpage

\clearpage

\begin{thebibliography}{}
\bibitem[Allard et al. 2001]{allard}Allard, F. et al. 2001, \apj, 556, 357
\bibitem[Barrado y Navascu\'{e}s \& Mart\'{i}n 2003]{bm03}Barrado y Navascu\'{e}s \& Mart\'{i}n 2003, \aj, 126, 2997
\bibitem[Barrado y Navascu\'{e}s et al. 2007]{b07}Barrado y Navascu\'{e}s, D. et al. 2007, \aap, 468, L5
\bibitem[Bate et al. 2003]{bate}Bate, M. R. et al. 2003, MNRAS, 339, 577
\bibitem[B\'{e}jar et al. 2008]{bejar}B\'{e}jar, V. J. S. et al. 2008, \apj, 673, L185
\bibitem[Bouvier et al. 1993]{bouv}Bouvier, J. et al. 1993, \aap, 409, 169
\bibitem[Caballero et al. 2006]{c06}Caballero, J. et al. 2006, \aap, 460, 635
\bibitem[Caballero 2009]{c09}Caballero, J. 2009, \aap, 507, 251
\bibitem[Carpenter 2001]{carp}Carpenter, J. 2001, \aj, 121, 2851
\bibitem[Carpenter et al. 2001]{carp01}Carpenter, J. et al. 2001, \aj, 121, 3160
\bibitem[Close, L. et al. 2007]{close}Close, L. et al. 2007, \apj, 660, 1492
\bibitem[Comer\'{o}n et al. 2004]{comeron}Comer\'{o}n, F. et al. 2004, \apj, 602, 298
\bibitem[Doppmann et al. 2005]{dopp}Doppmann, G. W. et al. 2005, \aj, 130, 1145
\bibitem[Ducourant et al. 2005]{ducou}Ducourant et al. 2005, \aap, 438, 769
\bibitem[Hawkins \& Veron 1993]{haw}Hawkins \& Veron 1993, MNRAS, 260, 202
\bibitem[Kama et al. 2009]{kama}Kama et al. 2009, \aap, 506, 1199
\bibitem[Kraus \& Hillenbrand 2009]{kh09}Kraus, A. L. \& Hillenbrand 2009, \apj, 703, 1511
\bibitem[Kusakabe et al. 2005]{kusakabe}Kusakabe et al. 2005, \apj, 632, L139 
\bibitem[Luhman 2004]{luh04}Luhman, K. L. 2004, \apj, 614, L398
\bibitem[Luhman \& Mamajek 2010]{lm10}Luhman, K. L. \& Mamajek, E. 2010, \apj, 716, L120
\bibitem[Luhman et al. 2007]{luh07}Luhman, K. L. et al. 2007, \apj, 659, 1629
\bibitem[Marley et al. 2002]{marley}Marley, M. et al. 2002, \apj, 568, 335
\bibitem[Mart\'{i}n \& Zapatero Osorio 2003]{mz03}Mart\'{i}n, E. L. \& Zapatero Osorio 2003, \apj, 593, L113
\bibitem[Minniti et al. 2010]{minniti}Minniti et al. 2010, New Astronomy, 15, 433
\bibitem[Park \& Kenyon 2002]{park} Park, S. \& Kenyon, S. 2002, \aj, 123, 3370
\bibitem[Persson et al. 1998]{perss}Persson, S. E. et al. 1998, \aj, 116, 2475
\bibitem[Scholz et al. 2005]{sch05}Scholz A. et al. 2005, \aap, 438, 675
\bibitem[Scholz et al. 2009]{sch09}Scholz A. et al. 2009, MNRAS, 398, 873
\bibitem[Terebey, S. et al. 1990]{t90} Terebey, S. et al. 1990, \apj, 362, L63 
\bibitem[Terebey, S. et al. 1998]{t98} Terebey, S. et al. 1998, \apj, 507, L71 (T98)
\bibitem[Terebey, S. et al. 2000]{t00} Terebey, S. et al. 2000, \aj, 119, 2341
\bibitem[Winn et al. 2006]{winn}Winn et al. 2006, \apj, 644, 510 
\bibitem[Zapatero Osorio et al. 2002]{zo02}Zapatero Osorio, M. R. et al. 2002, \aap, 384, 937
\bibitem[Zapatero Osorio et al. 2008]{zo08}Zapatero Osorio, M. R. et al. 2008, \aap, 477, 895
\end{thebibliography}
\end{document}